\documentstyle[prl,epsfig,aps]{revtex} 
\begin{document} 
 
\twocolumn[\hsize\textwidth\columnwidth\hsize\csname@twocolumnfalse\endcsname 
 
\title{Edwards measures for powders and glasses} 
 
\author{Alain Barrat$^1$, Jorge Kurchan$^2$,  
Vittorio Loreto$^3$  and Mauro Sellitto$^4$ }  
  
\address{   
$^1$ Laboratoire de Physique Th{\'e}orique  
\cite{umr}, B{\^a}timent 210, Universit{\'e} 
de Paris-Sud, 91405 Orsay Cedex, France \\ 
$^2$ P.M.M.H. Ecole Sup\'erieure de Physique et Chimie 
Industrielles, 10 rue Vauquelin 75231 Paris, France \\ 
$^3$  Universit\`a degli Studi di Roma ``La Sapienza'', 
Dipartimento di Fisica, P.le A. Moro 5, 00185 Rome, Italy \\ 
and INFM, Unit\`a di Roma 1\\ 
$^4$ Laboratoire de Physique 
             de l'\'Ecole Normale Sup\'erieure de Lyon,         
  46 All\'ee d'Italie, 69007 Lyon, France. 
} 
 
\date{\today} 
 
\maketitle 
\begin{abstract} 
 Can one construct a thermodynamics for compact, slowly moving  
powders and grains? 
A few years ago, Edwards proposed a possible step in this direction, 
raising the fascinating perspective that such systems  
have a statistical mechanics of their own, 
different from that of Maxwell, Boltzmann and Gibbs,  
allowing us to have some 
information while still ignoring dynamic details. 
 
Recent developments in the  theory of glasses   
have come to confirm  these ideas  within mean-field.  
In order to go beyond, we explicitly generate Edwards'  
measure in a model that, although schematic, is three 
dimensional. Comparison of  the results thus obtained with the 
irreversible  compaction data shows very good agreement.  
The present  framework immediately  suggests new experimental checks. 
 
\end{abstract} 
\twocolumn  
\vskip .5pc] 
\narrowtext 
 
The classical way to go from the microscopic dynamics to statistical 
mechanics proceeds in two steps: one first identifies a distribution that is 
left invariant by the dynamics ({\em e.g.} the microcanonical ensemble), and 
then assumes that this distribution will be reached by the system, under  
suitable conditions of 'ergodicity'. 
For granular systems this approach seems doomed from the 
outset: because energy is lost through internal friction, and 
gained by a non-thermal source such as tapping or shearing, 
the dynamical equations do not leave the microcanonical or any 
other known ensemble invariant. Moreover, the compaction 
dynamics is extremely slow and does not approach any  
stationary state on experimental time scales. This raises 
the question of characterizing the typical configurations 
or the region of phase space visited dynamically. 
 
The proposal of Edwards and collaborators~\cite{Sam,anita,Repo} is to 
use an alternative distribution for very gently vibrated or  
sheared granular systems, with the static situation as a limiting case. 
It may be summarized as follows: given a certain situation attained  
dynamically, physical observables are obtained by averaging over  
the usual equilibrium distribution at the corresponding volume,  
energy, etc. {\em but restricting the sum to the `blocked' configurations}  
defined as those in which every grain is unable to move. 
This definition leads immediately to an entropy  (in the glass  
literature a `complexity') $S_{edw}$, given by  the logarithm  
of the number of blocked configurations of given volume, energy, etc., 
and its corresponding  density $s_{edw}\equiv S_{edw}/N$. 
Associated with this entropy are the   state variables such as 
`compactivity' $X_{edw}^{-1}=\frac{\partial}{\partial V}S_{edw}(V)$ and 
`temperature' $T_{edw}^{-1}=\frac{\partial}{\partial E}S_{edw}(E)$. 
 
That configurations with low mobility should be relevant in a jammed 
situation is rather obvious, the strong assumption here is that,  
apart from the usual statistical weights, {\em all blocked configurations  
are treated as equivalent} --- any extra weight of dynamical origin  
that might distinguish them is disregarded.  
The purpose of this letter is to argue that this {\em `flatness'}  
assumption characterizing Edwards' distributions is neither capricious  
(it leads to correct predictions for the compaction dynamics of  
a given class of systems), nor obvious (it does not apply to other  
classes of systems).  
To do this we devise a method to count the  blocked configurations 
and compute averages over them. 
 
Let us briefly summarize the state of the art. 
A first clue comes from exploiting the analogy  
between the settling of grains and powders, 
as when we gently tap a jar with flour to make space for more,  
and the aging of glassy systems~\cite{Struik,Nagel,Nagel2}: in both cases, 
the system remains out of equilibrium on all accessible time-scales, 
and displays very slow relaxations.  
 
In the late eighties, Kirkpatrick et al.~\cite{KTW,KTW2} recognised that 
a class of mean-field models contains, although in a rather 
schematic way, the essentials of glassy phenomena. 
When the aging dynamics of these systems was solved analytically, 
a feature that emerged was the existence of a temperature $T_{dyn}$ 
for {\em all} the slow modes (corresponding to structural rearrangements) 
\cite{review,Cukupe}. 
For our purposes here, $T_{dyn}$ can be defined by comparing 
the random diffusion and the mobility between two widely 
separated times $t$ and $t_w$ of any particle or tracer 
in the aging glass.  
Surprisingly, one finds  in all cases an  Einstein 
relation  $\left\langle (r(t)-r(t_w))^2 \right\rangle 
 = T_{dyn}\frac{\delta \left\langle r(t) -r(t_w) \right\rangle }{\delta f}$, 
where $r$ is the position of the particle and
$f$ is a constant perturbing field. 
While in an equilibrium system the fluctuation-dissipation theorem  
guarantees that the role of $T_{dyn}$ is played by the  
thermodynamic temperature, 
the appearance of such a quantity out of equilibrium is by no  
means obvious. $ T_{dyn}$ is different from 
the external temperature, but it can be shown to have 
 all other properties defining a true temperature~\cite{Cukupe}. 
 
As it turned out, despite its very different origin, 
this temperature matches exactly Edwards' ideas:  
$T_{edw}$ and $T_{dyn}$ happen to coincide  
for mean-field glass models aging in contact  
with an almost zero temperature bath~\cite{remi,jamming,Theo,Frvi,Felix}.  
In fact, given the energy $E(t)$ at long times, the value of any other 
macroscopic observable is also given by  the flat average over all blocked 
configurations of energy $E(t)$. 
Within the same approximation, one can also treat systems that like 
granulars present a non-linear friction and different kinds of energy 
input, and the conclusions remain the same~\cite{jorge-trieste}.

A first partial  conclusion is then that Edwards' scenario 
is at the very least correct within mean-field schemes and for  
very weak  vibration or forcing. The problem that remains is  
to what extent it carries through to more realistic models.  
 
In this direction, there have been  recently  studies~\cite{inherent1}  
of Lennard-Jones glass formers from the perspective of the so-called  
`inherent structures' (a partition of the phase-space in terms of the  
blocked configurations~\cite{inherent}). 
In this context a `flat weight' assumption --- similar in spirit but  
not quite equivalent to Edwards' ---  also comes into question and is  
tested in various ways. Though there are  caveats\cite{enfants,Andrea},  
the results are encouraging. 
 
The path we  follow  is instead to construct the Edwards 
measure explicitly in the case of  
representative (non mean-field) systems, together with the corresponding 
entropy and expectation values of observables. 
We thus obtain results that are distinctly different  
from the equilibrium ones, 
and we can compare both sets  with those  
of the  irreversible compaction dynamics.

The first model we consider is the so-called Kob-Andersen (KA)  
model~\cite{KoAn} that, though very schematic, reproduces rather well  
several aspects of glasses~\cite{KuPeSe} and of granular  
compaction~\cite{SeAr}; most important, this model is non mean-field. 
A particle can move  
to a neighbouring empty site, on a three dimensional lattice, only if it  
has less than four neighbours in the initial and in the final position. 
(In these `hard particle' models the temperature is irrelevant, 
and we set it to one.) 
The dynamic  rule guarantees that the  equilibrium distribution is trivially  
simple since all the configurations of a given density are equally probable. 
However, at densities close to $\rho_{\rm g}$ ($\simeq .88$), the particle  
diffusion becomes extremely slow due to the kinetic constraints. 
In order to mimic a compaction (or aging) process 
without gravity, we simulate a `piston'  by freely creating and  
destroying particles only on the topmost layer with a  chemical  
potential $\mu$~\cite{KuPeSe}.

{\it i)} The dynamic measurements are taken  as follows: 
starting from low density, we perform a  slow compression 
by raising the chemical potential up to a high value $\mu=3$. 
Since  the equilibrium density at $\mu=3$ 
is much larger than the jamming density   $\rho_{\rm g}$, 
the system falls out of equilibrium and very slow compaction ensues. 
We record the density $\rho(t)$ and the spatial  
structure function $g_{dyn}(r,t)$ defined as the probability that 
two sites at distance $r$ are occupied.  
We also compute the  dynamic temperature $T_{dyn}$  
by comparing induced and spontaneous displacements. 
This is the set of observables we use for testing the different  
measures, which are obtained independently. 
 
{\it ii)} In the equilibrium measure all configurations  
(whether they are blocked or not)  
have equal weight. It is easy to obtain the exact  
equilibrium entropy density  per particle $s_{equil}(\rho)= 
-\rho \ln \rho - (1-\rho)\ln(1-\rho)$. 
Since $T=1$ as mentioned above, $\frac{ds_{equil}(\rho)}{d\rho}=-\mu$. 
The equilibrium structure factor is easily seen to be 
a constant $g_{equil}(r)=\rho^2$: indeed, one main advantage of this model 
is that it is particularly easy to compare small deviations  
from $g_{equil}(r)$, a  notoriously difficult task  in 
glassy systems. 
 
\begin{figure}[7] 
\epsfxsize=3.4in 
\centerline{\epsffile{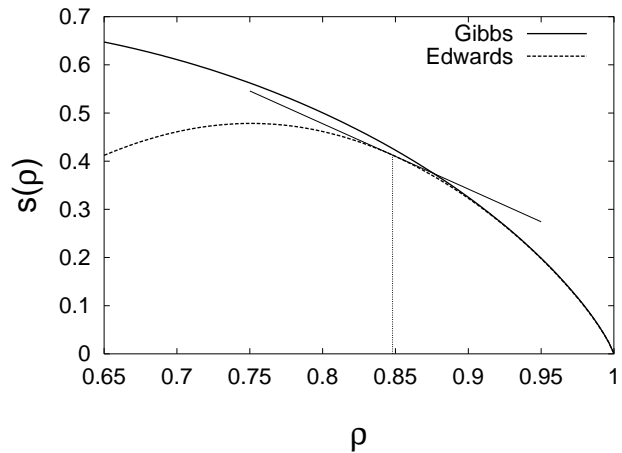}} 
\vspace{0.2cm} 
\caption{Gibbs and Edwards entropies per particle  
of the Kob-Andersen model vs. density. At high enough density  
the curves are indistinguishable, and join exactly only at $\rho=1$. The slope 
of the tangent to $s_{edw}(\rho)$ for a generic $\rho$ allows to extract 
$T_{edw}(\rho)$ from the relation 
$\frac{ds_{edw}}{d\rho}=\frac{1}{T_{edw}(\rho)} 
\frac{ds_{equil}}{d\rho}$. 
} 
\label{fig:entropia} 
\end{figure}

\begin{figure}[7] 
\epsfxsize=3.4in 
\centerline{\epsffile{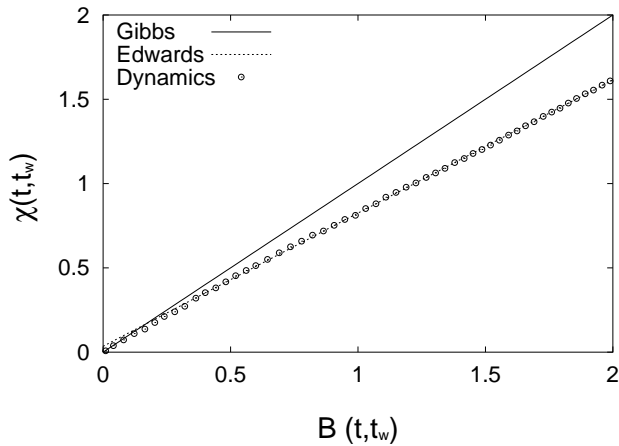}} 
\vspace{0.2cm} 
\caption{Einstein relation in the Kob-Andersen model: plot of  
the mobility  $\chi(t,t_w)$ vs. the mean-square displacement  
$B(t,t_w)$. The slope of the full straight line corresponds 
to the equilibrium temperature ($T=1$), and the slope of the dashed one to 
Edwards' prescription obtained from figure~\ref{fig:entropia} at 
$\rho(t_w)=0.848$.
} 
\label{fig:x} 
\end{figure}

{\it iii)} Finally, we obtain Edwards' measure results as follows: 
we introduce an `auxiliary model' in which particles have energy equal 
to one if the dynamic rule of the original model would allow them to
move, and to zero otherwise.
Performing  simulated annealing of the auxiliary model  at 
fixed number of particles is an  efficient way to sample over the 
configurations with vanishing  fraction of moving particles  the 
Edwards ensembles structure function $g_{edw}(r)$, and to obtain  
 (Figure~\ref{fig:entropia})
$S_{edw}(\rho)$ as the logarithm of the number of such configurations
by thermodynamic integration of the energy of the auxiliary model
with respect to its temperature. We then compute  
$T_{edw}^{-1}=-\frac{1}{\mu}\frac{ds_{edw}(\rho)}{d\rho}$. 
We are now in a position to compare the long-time results of the out of 
equilibrium dynamics {\em i)} with those obtained with  
measures {\em ii)} and {\em iii)}. 
Figure \ref{fig:x} 
shows a plot of the mobility 
$\chi(t,t_w)=\frac{1}{3N}\sum_{a=1}^{3} 
\sum_{k=1}^{N} \frac{\delta \left\langle (r_k^a(t)-r_k^a(t_w)) 
\right\rangle}{\delta f}$ 
{\em vs.} the mean square displacement
$B(t,t_w)=\frac{1}{3N}\sum_{a=1}^{3} 
\sum_{k=1}^{N} \left\langle (r_k^a(t)-r_k^a(t_w))^2 \right\rangle$, 
testing the existence of a dynamical temperature $T_{dyn}$,~\cite{Se}, 
in the compaction data 
($N$ is the number of particles and $a$ runs over 
the spatial dimensions).
The agreement between $T_{dyn}$ and the Edwards temperature $T_{edw}$, 
obtained from the blocked configurations as in Figure~\ref{fig:entropia}, 
is clearly excellent.  
\begin{figure}[7] 
\epsfxsize=3.4in 
\centerline{\epsffile{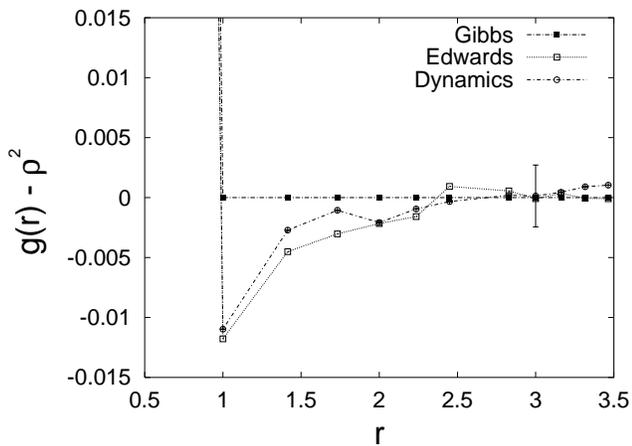}} 
\vspace{0.2cm} 
\caption{Structure functions $g(r) - \rho^2$ at 
density $\rho \simeq 0.87$ computed with the 
equilibrium, Edwards' and dynamical measure 
of the Kob-Andersen model. The three sets of data come from independent 
Monte-Carlo simulations. The dynamic structure function (circles) is obtained  
after  slow compression  raising the chemical potential continuously from 
$\mu=1$ to $\mu=3$ in $10^{6}$ Monte Carlo sweeps. 
The Edwards structure function (open squares) is obtained from the  
auxiliary model. Although the equilibrium 
value of $g(r)-\rho^2$ is exactly $0$, we also obtain it by a Monte-Carlo  
simulation (full squares) in order 
to show that the difference in the short distance behaviour is not an
artifact  
of the numerical simulation). The size of the typical error bar on dynamical  
data is shown at $r=3$. 
} 
\label{fig:g} 
\end{figure} 
In Figure \ref{fig:g} we plot {\em i)}
the long-time dynamic $g_{dyn}(r,t)$,
{\em ii)} the equilibrium  $g_{equil}(r)=\rho^2$,
and  {\em iii)} the Edwards $g_{edw}(r)$ structure
factors.
The agreement between {\em i)} and   {\em iii)} is good.
 
From the results shown so far, a picture emerges where the Edwards 
measure is able to correctly reproduce the sampling of the phase space 
generated by the out of equilibrium dynamics of this non mean-field model. 
We have found, however, that at short times or for excessively fast 
compressions, the quality of the agreement becomes worse, possibly due to 
heterogeneities. 
We refer to a longer, more technical paper for a discussion of these issues,
as well as a study of other models (in particular the so-called Tetris model
\cite{tetris} for which one recovers the same conclusions
as for the KA model) 
and more technical details on our numerical methods \cite{bkls}. 
 
As already mentioned, Edwards' construction can be inappropriate  
for certain models, even though they may have a logarithmically slow dynamics. 
As a representative  
example of this we consider the low temperature domain growth dynamics  
of a 3D Ising model in a weak random magnetic field, a model relevant to many 
physical problems~\cite{Nattermann}.  
At large times the domain walls are pinned by the field, and the  
dynamics proceeds by thermal activation. The mean energy
decreases slowly towards the ground state energy. In a large system,  
the long-time configurations are made of domains  
of `up' and `down' spins having similar volumes, 
the global magnetization being zero. 
This is quite different from the equilibrium configurations 
at the same energy, which are instead magnetized (since the energy is near
the ground state energy). 
 
The question in the present context is therefore whether a  
long-time configuration of (low) energy $E_0$
is well reproduced by the {\em typical} `blocked'  
configuration of the {\it same} energy. 
By simulating the corresponding `auxiliary' model, (with auxiliary energy 
equal to the number of spins not aligned with their local field,
i.e. to the number of 'mobile' spins),  
we have checked that this is not the case: 
the blocked configurations consituting Edwards' distribution at energy $E_0$
are also {\em magnetised}. Therefore, neither Gibbs' 
nor Edwards' distributions describe the typical configurations 
obtained dynamically. 
 
\begin{figure}[7] 
\epsfxsize=3.4in 
\centerline{\epsffile{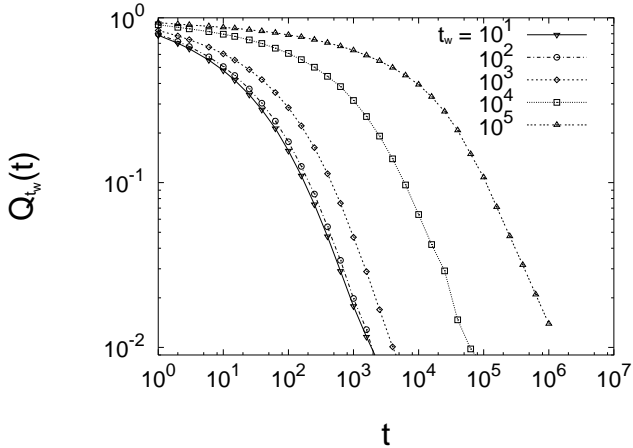}} 
\vspace{0.1cm} 
\caption{Mean overlap $Q_{t_w}(t)$ 
between two clones in the KA model: the two clones are separated at $t_w$
and evolve subsequently with different noises. $Q_{t_w}(t)$ always decreases
to zero (the slower the larger $t_w$), showing that the clones
always diverge.
} 
\label{fig:clone} 
\end{figure}

When is then the flatness assumption characterizing Edwards' argument 
justified? A natural criterion, suggested by glass 
theory~\cite{baldassarri,cude,bamebu,Frvi}, consists in 
studying how a system explores its phase space, i.e. its 
`chaoticity' properties. After aging for a time $t_w$,  
two copies (clones) are made of the system, 
and allowed to evolve subsequently 
with different realisations of the randomness {\em in the updating procedure}. 
We have checked that in the KA model the  
two clones always diverge (the slower the larger 
$t_w$, see figure \ref{fig:clone}), while for the 3D Random Field 
Ising model they do not. It is thus tempting 
to conjecture that this form of chaoticity is a necessary condition 
to have   flat  statistical weights for the blocked  
configurations. 
Note that for this criterion to make sense, it should always be applied 
at non-zero (though weak) tapping or shearing. 
 The condition of chaoticity is however not sufficient: Bouchaud's
 `trap model' \cite{review} is chaotic but its fluctuation-dissipation
 properties are not directly related to the density of states.
To summarize, our study suggests that the proposal made by Edwards 
does indeed make sense and opens a door towards a statistical  
(thermodynamic) 
description of compact granular matter under very weak driving. 
In order to generalise these ideas to stronger forcing, lower chemical  
potential or higher temperatures (as required to analyse the 
experiments in~\cite{sid}), one has to learn how to go from the   
concept of `blocked configuration' to that of `metastable state',  
and this requires other tools~\cite{giulio}. The inherent structure 
 construction could provide a practical shortcut. 
 
The present setting of the problem immediately suggests experiments to check  
these ideas, e.g. by studying diffusion and mobility of tracer particles 
within driven granular media. 
 
Finally, let us note that even in the simplest cases, the 
correspondence between Edwards' distribution and long-time dynamics 
is at best checked but does not follow from any principle. 
The situation is thus as if one would have checked that  
the microcanonical distribution gives good results for gases, 
without knowing Liouville's  theorem that proves that such a distribution is 
indeed left invariant by the equations of motion.  
Such more refined arguments would be very welcome. 

\bigskip
\bigskip
  
ACKNOWLEDGEMENTS:  MS is supported by a Marie Curie fellowship of the 
European Commission (contract ERBFMBICT983561). This work has also been 
partially supported from the European Network-Fractals under contract 
No. FMRXCT980183.

\end{document}